\documentclass[preprint, prd]{revtex4}
\usepackage{hyperref}

\usepackage{amsmath,amssymb,amsfonts}
\usepackage{graphicx}
\newcommand{\abs}[1]{\left\vert #1 \right\vert}

\newcommand{\del}[2]{\frac{\partial#1}{\partial #2}}
\newcommand{\dif}[2]{\frac{d#1}{d#2}}

\newcommand{\m}[1]{\mathbf{#1}}

\begin{document}

\title{Anti-telephones in transformation optics:\protect\\metamaterials with closed null geodesics}
%title{Galadriel's Mirror: \protect\\ anti-telephones from transformation optics}
\date{February 20, 2015}
%\received{}

\author{S. Reece \surname{Boston}}
\email{rboston@physast.uga.edu}
%% AFFILIATION IS NECESSARY, MUST HAVE FULL ADDRESS
\affiliation{Department of Physics and Astronomy, University of Georgia, Athens, GA 30602}

\begin{abstract}
We apply the methods of transformation optics to theoretical descriptions of spacetimes that support closed null geodesic curves.  The metric used is based on frame dragging spacetimes, such as the van Stockum dust or the Kerr black hole.  Through transformation optics, this metric is analogous to a material that in theory should allow for communication between past and future.  Presented herein is a derivation and description of the spacetime and the resulting permeability, permittivity, and magneto-electric couplings that a material would need in order for light in the material to follow closed null geodesics.  We also address the paradoxical implications of such a material, and demonstrate why such a material would not actually result in a violation of causality.  A full derivation of the Plebanski equations is also included.
\end{abstract}

%% I NEED PACS CODES
%differential geometry

\maketitle

%%%%%%%%%%%%%%%%%%%%%%%%%%%%%%%%%%%%%%%%%%%%%%%%%%%
%			INTRODUCTION
%%%%%%%%%%%%%%%%%%%%%%%%%%%%%%%%%%%%%%%%%%%%%%%%%%%
\section{Introduction\label{Introduction}}

Since the first papers by Pendry  \cite{Pendry1} and Leonhardt \cite{Leonhardt1}, the subject of invisibility cloaks has garnered much attention in the literature.  For the design of these cloaks, the recent field of transformation optics (TO) was developed.  Though TO bestows on the scientist near-unlimited control of the movement of light, most research efforts have been directed to perfecting the invisibility devices that initially drew attention to the field.

Tranformation optics uses the coordinate-invariance of Maxwell's equations to set up an analogy between Maxwell's equations in curved vacuum spacetime to Maxwell's equations in a flat spacetime with a particular medium.  Therefore, so long as we have a mathematical description of a curved spacetime, an analogous material can be constructed within which light will behave similarly to the curved spacetime.  For a fuller introduction to the subject of transformation optics, the reader is referred to the excellent paper by Leonhardt and Philbin \cite{LeonhardtTO}.

Due to the direct parallels between TO and general relativity (GR), some interest has been directed towards the design of materials that would simulate various models of spacetime, hopefully allowing astronomers to study systems such as black holes in the laboratory.  Proposals for materials mimicking the effects of DeSitter space \cite{DeSitter}, Schwarzschild black holes \cite{Schwarzschild}, Kerr black holes \cite{Thompson1}, spatial wormholes \cite{Greenleaf1}, Alcubierre warp drive geometries \cite{Warpdrive},  and so-called optical black holes \cite{OpticalBH,celestialmechanicsinTO}  have been put forward.  The possibilities for materials are nearly endless, and any kind of spacetime geometry, no matter how bizarre --- including those which do not solve the Einstein equations --- can, at least in theory, be modeled using the formalism of TO.

Herein, we propose a material that will use transformation optics to simulate one of the more imaginative and hotly-debated aspects of curved spacetimes; namely, the possibility of time travel to the past along closed causal curves.   In the literature, most interest in time travel metrics has been directed toward curves that allow matter to move back in time, often called closed time-like curves.  In TO, we are interested in the movement of light, which in GR is known to follow null geodesics.  Of particular interest is the case of closed null geodesics (CNGs), which would form a continuous loop in lab time from future to past.  In Section \ref{Relativity} we construct a metric that produces CNGs, without regard to the usual physical constraints of the energy conditions.

Using the same procedure that led to the invisibility cloak design and the dielectric black hole design, we use the metric tensor of our spacetime supporting CNGs to generate material parameter tensors (MPs), namely permittivity $\epsilon$, permeability $\mu$, and magneto-electric tensors $\gamma_1,\gamma_2$, whose effect on the fields is expressed by
\begin{equation}
\vec{D} = \epsilon \vec{E} + \gamma_1\vec{H}, \quad \vec{B} = \mu \vec{H} + \gamma_2\vec{E}.
\end{equation}
We will show that within this resultant material there exist CNGs that can span a finite lab time difference $T$, allowing signaling from the future to the past along the CNGs.  This would, in theory, allow information from the future to have an impact on the past.

In Section \ref{Spacetime} we discuss and derive the Plebanski equations \cite{Plebanski}, which relate curved spacetimes to material parameters, following the Minkowski formalism of electromagnetism.  After this, in Section \ref{Relativity} we propose a simplified spacetime metric that should produce such curves.  In Section \ref{Galadriel} we combine the results of the previous sections and explicitly solve for the MPs based on the formalism presented.  We also provide a sketch showing how such a material could be used to communicate with the past.  Finally, in Section \ref{Limitations}, we point out technical limitations which should impede the actual functioning of any such device as that proposed here.

%%%%%%%%%%%%%%%%%%%%%%%%%%%%%%%%%%%%%%%%%%%%%%%%%%%
%			TO in SPACETIME
%%%%%%%%%%%%%%%%%%%%%%%%%%%%%%%%%%%%%%%%%%%%%%%%%%%
\section{Transformation Optics in Spacetime\label{Spacetime}}

We begin with the field-strength tensor $\m{F} = d\m{A}$, with Cartesian components
\begin{equation}\label{fieldtensor1}
	F_{\mu\nu} = \left( \begin{matrix} 0& -E_x & -E_y & -E_z\\ E_x & 0 & B_z & -B_y\\ E_y & -B_z & 0 & B_x\\ E_z & B_y & -B_x & 0  \end{matrix} \right),
\end{equation}
and its dual tensor $\m{G}=\star\m{F}$, with Cartesian components
\begin{equation}\label{fieldtensor2}
	G_{\mu\nu}=\left(\begin{matrix} 0& H_x & H_y & H_z\\ -H_x & 0 & D_z & -D_y\\ -H_y & -D_z & 0 & D_x\\- H_z & D_y & -D_x & 0 \end{matrix}\right),
\end{equation}
both given in Minkowski spacetime using $c=1$ and the (-+++) sign convention for the metric.

In terms of these two tensors, the four Maxwell equations simplify to the two equations
\begin{subequations}
\begin{align}
	d\m{F} = 0 &\Longleftrightarrow \nabla_{[\alpha}F_{\beta\gamma]} = 0\\
	d\m{G} = \m{J} &\Longleftrightarrow \nabla_{[\alpha}G_{\beta\gamma]} = \sqrt{\abs{g}}\epsilon_{\alpha\beta\gamma\delta}j^\delta
\end{align}
\end{subequations}
where $g = \det(\m{g})$ and $\m{g}$ is the metric tensor of spacetime.

The defining equation $\m{G} = \star\m{F}$ states that $\m{G}$ is the Hodge dual of $\m{F}$, where $\star$ is the Hodge star operator, whose effect on $\m{F}$ can be expressed in component form by
\begin{equation}\label{constitutive}
	\m{G} = \star\m{F} \Longleftrightarrow G_{\alpha\beta} = \frac{1}{2}\sqrt{\abs{g}}\epsilon_{\alpha\beta\mu\nu}g^{\mu\lambda}g^{\nu\kappa}F_{\lambda\kappa}.
\end{equation}

From here, Thompson et al. \cite{Thompson2} go on to derive a covariant expression of transformation optics in spacetime.  They propose a slightly more general relationship, $\m{G} = \boldsymbol{\chi}(\star\m{F})$, where $\boldsymbol{\chi}$ is an antisymmetric four-tensor that is meant to contain all information about media in the spacetime.  They then apply coordinate transformations to express the effect of a curved spacetime with metric $\m{g}$ in terms of a flat spacetime with a medium given by $\boldsymbol{\chi}$.  Their result proves TO to be a covariant theory appiclable to any coordinate system.

Having noted their result, and the concomitant assurance that TO works in any spacetime, we take a slightly simpler approach, following that of Plebanski \cite{Plebanski}.  We will work only in Cartesian coordinates, and assume our medium to be stationary relative to the laboratory coordinates.  Let us start with \eqref{constitutive},
\begin{equation}
	G_{\mu\nu} = (\star\m{F})_{\mu\nu} = \frac{1}{2}\sqrt{\abs{g}} \epsilon_{\mu\nu\alpha\beta}g^{\alpha\lambda}g^{\beta\kappa} F_{\lambda\kappa}
\end{equation}
From the forms given by \eqref{fieldtensor1} and \eqref{fieldtensor2}, clearly $E_a = F_{a0}$, $H_a=G_{0a}$, $D_a=\frac{1}{2}\epsilon_{abc}G_{bc}$, and $F_{ab}=\epsilon_{abc}B_c$.  Let us first consider specifically the $D_1=G_{23}$ component.  Then
\begin{eqnarray*}
	D_1 &=& \frac{\sqrt{\abs{g}}}{2}\epsilon_{23\alpha\beta}g^{\alpha\lambda}g^{\beta\kappa}F_{\lambda\kappa}\\
	&=& \frac{\sqrt{\abs{g}}}{2}\epsilon_{23\alpha\beta}\left[g^{\alpha0}g^{\beta a}F_{0a}+g^{\alpha a}g^{\beta 0} F_{a0} + g^{\alpha a}g^{\beta b}F_{ab}  \right]\\
	&=&\frac{\sqrt{\abs{g}}}{2}\left[ 2g^{0a}g^{10}F_{a0}-2g^{1a}g^{00}F_{a0}+2g^{0a}g^{1b}F_{ab}  \right]\\
	&=& \sqrt{\abs{g}} (g^{10}g^{0a}-g^{00}g^{1a})E_a + \sqrt{\abs{g}}g^{0a}g^{1b}\epsilon_{abc}B_c.
\end{eqnarray*}
Performing the same evaluation on the other two components, we find
\begin{equation}
	D_a = \sqrt{\abs{g}}(g^{a0}g^{b0}-g^{00}g^{ab})E_b + \sqrt{\abs{g}}g^{ad}g^{0c}\epsilon_{dcb}B_b.
\end{equation}
Next we look at $H_1=G_{01}$.  Here
\begin{eqnarray*}
	H_1 &=& \frac{\sqrt{\abs{g}}}{2}\epsilon_{01\alpha\beta}g^{\alpha\lambda}g^{\beta\kappa}F_{\lambda\kappa}\\
	&=& \frac{\sqrt{\abs{g}}}{2}\epsilon_{1ab}\left[  g^{a0}g^{bc}F_{0c}+g^{ac}g^{b0}F_{c0}+g^{ac}g^{bd}F_{cd}   \right]\\
	&=&\sqrt{\abs{g}}\epsilon_{1ab}g^{a0}g^{bc}E_c + \frac{1}{2}\sqrt{\abs{g}}\epsilon_{1ab}g^{ac}g^{bd}\epsilon_{cde}B_e,
\end{eqnarray*}
with the general result
\begin{equation}
	H_a = \sqrt{\abs{g}}\epsilon_{abc}g^{b0}g^{cd}E_d + \frac{\sqrt{\abs{g}}}{2}\epsilon_{abc}g^{bd}g^{ce}\epsilon_{def}B_f
\end{equation}
These two equations can be written more simply as
\begin{equation}
	D_a = e^{ab}E_b+f^{ab}B_b, \qquad H_a = h^{ab}B_{b} + k^{ab}E_b,
\end{equation}
and, upon rearrangement and simplification to the standard form, we arrive at the result first cited by Plebanski,
\begin{subequations}\label{plebanskiequation}
\begin{align}
	D_a &= -\frac{\sqrt{\abs{g}}}{g_{00}}g^{ab}E_b + \epsilon_{abc}\frac{g_{0b}}{g_{00}}H_c\\
	B_a &= -\frac{\sqrt{\abs{g}}}{g_{00}}g^{ab}H_b  - \epsilon_{abc}\frac{g_{0b}}{g_{00}}E_c.
\end{align}
\end{subequations}
In terms of the MPs, this gives 
\begin{subequations}\label{tensorequation}
\begin{align}
	\epsilon^{ab} =\:\: \mu^{ab}\;\: &= -\frac{\sqrt{\abs{g}}}{g_{00}}g^{ab}\\
	\gamma^{ab}_1=(\gamma^{T}_2)^{ab} &= \epsilon_{acb}\frac{g_{0c}}{g_{00}},
\end{align}
\end{subequations}
where
\begin{equation}
	D_a = \epsilon^{ab}E_b + \gamma_{1}^{ab} H_b, \qquad B_a = \mu^{ab}H_n + \gamma_2^{ab} E_b.
\end{equation}
Thus the curved spacetime of $g_{\alpha\beta}$ is equivalent to a flat spacetime with MPs $\epsilon^{ab}, \mu^{ab}, \gamma_1^{ab}, \gamma_2^{ab}.$

As Thompson and Plebanski both note \cite{Thompson2, Plebanski}, the above equations do not conserve index type, nor are they covariant, and they are only applicable to stationary media within a locally flat lab frame in Cartesian coordinates.  Despite this, the above equations are completely equivalent to the more general, covariant approach in \cite{Thompson2} for the case of a stationary medium, and can be extended to the covariant equations. These equations should therefore be applicable to a medium intended to emulate the effects of a curved spacetime.

%%%%%%%%%%%%%%%%%%%%%%%%%%%%%%%%%%%%%%%%%%%%%%%%%%%
%			CLOSED TIME-LIKE LOOPS
%%%%%%%%%%%%%%%%%%%%%%%%%%%%%%%%%%%%%%%%%%%%%%%%%%%
\section{Constructing a Metric with Closed Null Geodesics\label{Relativity}}

There has been much debate in scientific circles about the possibility of time travel to the past.  Forward time travel is, of course, trivially simple to achieve; it is the reverse situation, however, that gives us such trouble.  Most proposals require either particular models of the entire universe that are empirically false (for instance the G\"odel metric \cite{Godel}), or else highly idealized systems that cannot be physically realized, such as the negative energy densities of wormholes or the infinite rotating systems of van Stockum spacetimes \cite{vanStockum}.  For this reason, many physicists are comfortable dismissing the predicted causality violations in these contrived spacetimes as purely mathematical and unphysical --- as they say, ``garbage in/garbage out.''  More damning, Stephen Hawking has proposed a mechanism dubbed the Chronology Protection Conjecture \cite{Hawking}, whereby fields approaching CTCs (should any exist) are shown to be unstable --- leading to divergent stress-energies --- and hence cannot be supported, prohibiting time travel in this manner.  For a fuller discussion of the possibilities of time travel in a general relativistic framework, the reader is referred to a lecture by Thorne \cite{KipThorne} on the topic of the possibility of CTCs in GR.

Whether such a spacetime can or cannot be realized physically through various arrangements of stress-energy as in GR is not of interest of this present work; we make our curved spacetimes with metamaterials, not stress-energy densities.  We are interested, however, that a spacetime with CNGs can be described mathematically in terms of a metric tensor.

As in equation \eqref{plebanskiequation} above, if we have a spacetime with a metric $g_{\alpha\beta}$, it is possible to use this metric to calculate a related material with MPs $\epsilon^{ab}, \mu^{ab}, \gamma_1^{ab}, \gamma_2^{ab}$ so that light inside the material emulates light in the curved spacetime.   We will now explicitly construct a metric that allows for CNGs.  Our calculation is intended as a proof-of-concept, and hence we will keep the result as theoretically simple as possible.

We begin with a simple metric with two unspecified functions:
\begin{equation}\label{metricform}
	ds^2 = -dt^2 + dr^2 + B d\phi^2 + dz^2 + 2F d\phi dt
\end{equation}
where $c=1$ and $B$ and $F$ are, in general, functions of $r$ and $\phi$ only.  We will use the most general case for now, and apply some reasonable restrictions later.  The metric tensor and its inverse are then
\begin{equation}\label{metrictensor}
	g_{\alpha\beta} = \left( \begin{matrix} -1 &0&F&0\\ 0&1&0&0\\F&0&B&0\\0&0&0&1\end{matrix}\right), \qquad
	g^{\alpha\beta} = \left(\begin{matrix} \frac{-B}{B+F^2}&0&\frac{F}{B+f^2}&0\\0&1&0&0\\\frac{F}{B+F^2}&0&\frac{1}{B+F^2}&0\\0&0&0&1\end{matrix}\right).
\end{equation}
The condition for a curve to be a null curve is
\begin{equation}\label{nullequation}
	0 = ds^2 = -dt^2 + dr^2 + Bd\phi^2 + dz^2 + 2F dt d\phi.
\end{equation}
In addition to the null condition, a null geodesic also satisfies the geodesic equations
\begin{equation}\label{geodesicequation}
	\dif{^2x^\alpha}{\lambda^2} + \Gamma^\alpha_{\beta\gamma}\dif{x^\beta}{\lambda}\dif{x^\gamma}{\lambda} = 0,
\end{equation}
for any affine parameter $\lambda$.  This requires knowledge of the Christoffel symbols, which we can find using the formula
\begin{equation}\label{christoffelformula}
	\Gamma^\alpha_{\beta\gamma}  = \frac{1}{2}g^{\alpha\delta}(g_{\delta\beta,\gamma} + g_{\delta\gamma,\beta} - g_{\beta\gamma,\delta}).
\end{equation}
Solving, all vanish, except for
\begin{eqnarray*}
	\Gamma^t_{tr} =  \frac{1}{2}\frac{F}{B+F^2}\del{F}{r} ,&& \Gamma^\phi_{tr} \frac{1}{2}\frac{1}{B+F^2}\del{F}{r}\\
	\Gamma^r_{t\phi} = - \frac{1}{2}\del{F}{r}, && \Gamma_{\phi\phi}^r = -\frac{1}{2} \del{B}{r}\\
	\Gamma^t_{r\phi} = \frac{1}{2}\frac{F\del{B}{r} - B\del{F}{r}}{B+F^2}, && \Gamma_{r\phi}^\phi = \frac{1}{2}\frac{F\del{F}{r} + \del{B}{r}}{B+F^2}\\
	\Gamma^t_{\phi\phi} =  \frac{-B\del{F}{\phi} + \frac{1}{2}F\del{B}{\phi}}{B+F^2}, && \Gamma^\phi_{\phi\phi} = \frac{F\del{F}{\phi} + \frac{1}{2}\del{B}{\phi}}{B+F^2}
\end{eqnarray*}

We now impose a specific form for our curve, having constant radius and height; that is, ${t=t(\phi)},$ ${r=const}, {z=const}$, and we are using $\phi$ to parameterize.  To enforce closure of the curve,  we require that ${t(\phi)=t(\phi+2\pi)}$.  For simplicity, we will use $u(\phi)=\dif{t}{\phi}$ in all future equations.

There are two first integrals of note.  The null condition \eqref{nullequation} becomes
\begin{equation}
	\label{null}0 = -u^2 + 2Fu + B.
\end{equation}
Also, since our spacetime has no time dependence, we have the Killing equation (see \cite{MTW}) with Killing vector ${\xi^\alpha=(1,0,0,0)},$ which leads to
\begin{equation}
	\label{kt} -K = g_{\alpha\beta}\xi^\alpha u^\beta = -u + F,
\end{equation}
for constant $K$.  Further, the geodesic equations give us
\begin{eqnarray}
	\label{tG}0 &=& \dif{u}{\phi}  -\frac{B}{B+F^2}\del{F}{\phi} + \frac{1}{2}\frac{F}{B+F^2}\del{B}{\phi} \\
	\label{phiG}0 &=&   2F\del{F}{\phi} + \del{B}{\phi}\\
	\label{rG}0 &=&  -2 \del{F}{r} u - \del{B}{r}.
\end{eqnarray}
We now attempt to solve this system of coupled equations for $F$ and $B$.

Consider \eqref{kt}; here $u=u(\phi)$ and $F = F(r,\phi)$.  However, the two added together equals a constant, $K$.  Therefore, it must be the case that ${F(r,\phi) = f(\phi)}.$  We can then cross out all $\del{F}{r}$ terms in equations \eqref{tG}-\eqref{rG}.  Now let us combine \eqref{null} with \eqref{kt}; this gives us
\begin{equation}
	B = K^2 - f^2,
\end{equation}
which likewise implies that $B(r,\phi) = b(\phi)$, and we can remove all $\del{B}{r}$ terms.  Note that in this case, with ${K^2 = f^2+b}$, equations \eqref{phiG} and \eqref{rG} are trivially satisfied.  The last step is to check \eqref{tG}.  Note that in terms of $u$,
\begin{eqnarray}
	\label{metricF}	F(r,\phi) = f(\phi) &=& u -K\\
	\label{metricB}	B(r,\phi) = b(\phi) &=& 2Ku - u^2.
\end{eqnarray}
From this,  we find,
\begin{equation*}
	\del{F}{\phi} = \dif{u}{\phi}, \quad \del{B}{\phi} = 2K\dif{u}{\phi} - 2u\dif{u}{\phi} = -2f\dif{u}{\phi}
\end{equation*}
such that
\begin{eqnarray*}
 \dif{u}{\phi} - \frac{b}{K^2} \del{F}{\phi} + \frac{1}{2} \frac{f}{K^2}\del{B}{\phi} &=& \dif{u}{\phi} - \frac{b}{K^2} \dif{u}{\phi} - \frac{f^2}{K^2}\dif{b}{\phi}\\
 	&=& \dif{u}{\phi}\left(1 - \frac{b}{K^2} - \frac{f^2}{K^2}  \right)\\
	&=& \dif{u}{\phi} \left[ 1 - \frac{K^2}{K^2}\right] = 0.
\end{eqnarray*}
Therefore all four geodesic equations are satisfied for a curve ${u^\alpha = (u(\phi), 0, 1, 0)},$ in a spacetime with metric components defined as in \eqref{metricB}, \eqref{metricF}.  Note that here $u(\phi)$ is unspecified.  We are free to pick $u$ however we'd like, subject to the restriction that ${u(\phi) = u(\phi+2\pi)}$.  Here we will choose the path,
\begin{eqnarray}\label{timephi}
	t(\phi) &=& T \sin^2 \frac{\phi}{2}\\
	u(\phi) &=& h(\phi) = T \sin \phi,
\end{eqnarray}
which will loop up by $T$ in time, before looping back to the origin.

We can write the metric, in terms of $h(\phi) = T\sin\phi$, as
\begin{equation}
	\label{themetric} g_{\alpha\beta} = \left( \begin{matrix} -1 &0&h-K&0\\ 0&1&0&0\\h-K&0&h(2K-h)&0\\0&0&0&1\end{matrix}\right).
\end{equation}

Suppose we consider a new geodesic, parameterized by affine parameter $\lambda$ with four-velocity $u^\alpha = (\dot{t}, \dot{r},\dot{\phi},\dot{z}).$  This geodesic satisfies the geodesic equations, which are now simplified to
\begin{eqnarray}
	0 &=& \ddot{t} + \frac{\dot{\phi}^2}{K^2}\left(\frac{1}{2}f\del{b}{\phi} - b\del{f}{\phi}\right)\\
	0 &=& \ddot{r}\\
	0 &=& \ddot{\phi} + \frac{\dot{\phi^2}}{K^2}\left(f\del{f}{\phi} + \frac{1}{2}\del{b}{\phi}\right)\\
	0 &=& \ddot{z},
\end{eqnarray}
where $f(\phi) = h(\phi)-K$ and $b(\phi) = 2Kh(\phi)-h(\phi)^2.$  These solve as
\begin{eqnarray}
	\dot{t}(\lambda) &=& \dot{\phi}_o  h(\phi)\\
	r(\lambda) &=& \dot{r}_o\lambda + r_o\\
	\phi(\lambda) &=& \dot{\phi}_o\lambda + \phi_o\\
	z(\lambda) &=& \dot{z}_o\lambda + z_o,
\end{eqnarray}
giving the requirements for a geodesic in this spacetime.  If we require the geodesic to be null, and also to satisfy the Killing equations, then we find $\dot{r}_o=\dot{z}_o = 0$, and $\dot{t}(\phi) = \dot{\phi}_o h(\phi)$, returning our desired curve, up to choice of parameterization in $\dot{\phi}_o$.

We have thus proved that our spacetime permits a closed null geodesic.  It is crucial that $K\neq0$, for then our metric is singular.  We will now move on to combining this equation with the Plebanski equations, to solve for the analogous medium.

%%%%%%%%%%%%%%%%%%%%%%%%%%%%%%%%%%%%%%%%%%%%%%%%%%%
%			GALADRIEL'S MIRROR
%%%%%%%%%%%%%%%%%%%%%%%%%%%%%%%%%%%%%%%%%%%%%%%%%%%
\section{The Anti-Telephonic Medium \label{Galadriel}}

The metric above has been shown to support closed null geodesics --- paths along which light is known to move backwards relative to lab time.  Transformation optics allows us to use the spacetime metric \eqref{themetric} to formulate a material within which light should exhibit the same behavior as in the curved spacetime above; namely, light in the material should also move in CNGs.   We propose a device that uses this material to communicate with the past or the future, which we fancifully refer to as ``Galadriel's Mirror''\cite{Tolkien}.

Our metric tensor in Cartesian components is
\begin{equation}
g_{\alpha\beta} = \left(\begin{matrix}-1&-\frac{fy}{r^2}&\frac{fx}{r^2}&0\\
		-\frac{fy}{r^2} &1-(1-\frac{b}{r^2})\frac{y^2}{r^2} & (1-\frac{b}{r^2})\frac{xy}{r^2} & 0\\
		\frac{fx}{r^2} & (1-\frac{b}{r^2})\frac{xy}{r^2} & 1 - (1-\frac{b}{r^2})\frac{x^2}{r^2} &0\\
		0 & 0 & 0 & 1   \end{matrix}\right),
\end{equation}
still using $c=1$.  We calculate the MPs as in \eqref{tensorequation}.  These we find to be
\begin{subequations}\label{galadrieltensors}
\begin{align}
\epsilon^{ab} = \mu^{ab} &=\left(\begin{matrix} \frac{r}{K}\frac{y^2}{r^2} + \frac{K}{r}\frac{x^2}{r^2} & (\frac{K}{r}-\frac{r}{K})\frac{xy}{r^2} & 0\\
			(\frac{K}{r}-\frac{r}{K})\frac{xy}{r^2} & \frac{r}{K}\frac{x^2}{r^2}+\frac{K}{r}\frac{y^2}{r^2} & 0\\
			0 & 0 & \frac{K}{r}  \end{matrix}\right)\\
\gamma^{ab}_1= (\gamma_2^T)^{ab}&=  \left(\begin{matrix} 0 & 0 & \frac{f}{r}\frac{x}{r}\\0 & 0 & \frac{f}{r}\frac{y}{r}\\ - \frac{f}{r}\frac{x}{r} & -\frac{f}{r}\frac{y}{r} & 0\end{matrix}\right)
\end{align}
\end{subequations}
listed in Cartesian components.  These material parameters, if realized, should lead to the formation of closed causal curves, as discussed already in Section \ref{Relativity}.

We now sketch a rough picture of how the Mirror might be used to send anti-telephonic signals.  

We suppose, first of all, some cylinder with MPs as above.  The cylinder need not be solid, to avoid the singularities when $r=0$.  We also assume that $r\leq K$, which will let $\epsilon, \mu$ retain non-negative values.  Outside of the cylinder, there needs to be a method of shunting light into and out of the CNGs; this might be accomplished by means of fiberoptic cables pointed tangentially to the cylinder.  At any time between $t=t_o$ and $t=t_f$ (with $T=t_f-t_o$), the light can be extracted at a suitable angular position around the Mirror, according to $t=t(\phi)$ as in \eqref{timephi}.  By whatever means light is inserted and extracted, it will be important that the future recipients receive the message it encodes sometime before $t=t_f$, to give them time to respond.

\begin{figure}
\includegraphics[width=\columnwidth]{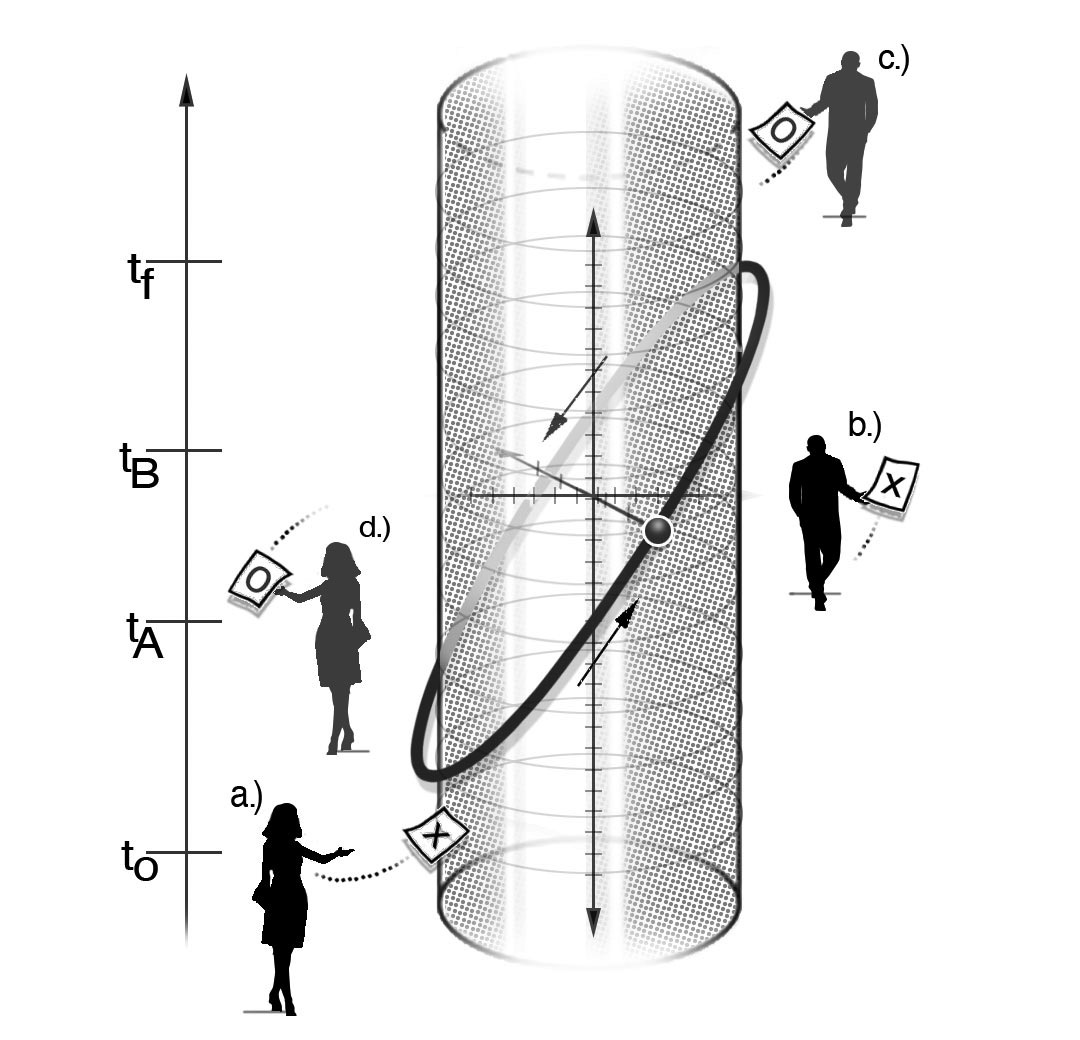}
\caption{Alice and Bob, separated by total time $T=t_f-t_o$, communicate using the Mirror.  The axis of the cylinder is along the time axis.  (a.) Alice encodes her message (``X'') at time $t=t_o$ and places it in the Mirror.  (d.) Alice receives Bob's message (``O'') at time $t=t_A$.  (b.) Bob receives Alice's message of ``X'' at time $t=t_B$.  (c.) At any time before $t=t_f$, Bob replies to Alice with his own message of ``O''.  After $t_f$ in the lab frame of reference, the light in the Mirror vanishes.}
\label{aliceandbob}
\end{figure}

In Figure \ref{aliceandbob} we provide a visual illustration for how information from the future can be sent to the past.  Here, Alice and Bob are communicating by light signals propagating along a CNG within the Mirror.  The optical spacetime geometry of the Mirror allows for Bob's message to reach in to the past, enabling Alice to know the future.  This presents an apparent violation of causality.  We now look at possible solutions to this quandary.

%%%%%%%%%%%%%%%%%%%%%%%%%%%%%%%%%%%%%%%%%%%%%%%%%%%
%			LIMITATIONS
%%%%%%%%%%%%%%%%%%%%%%%%%%%%%%%%%%%%%%%%%%%%%%%%%%%
\section{Limitations to Causality Violation\label{Limitations}}

The preservation of causality is essential in the classical understanding of physics.  Therefore, it is prudent to examine a number of possible mechanisms that will prevent a device like Galadriel's Mirror from working as outlined above.

Within the context of general relativity, there exist situations where the standard understanding of causality can become muddled.  As an example, supposing a laboratory frame moving along a closed time-like curve (CTC), the causal order inside the laboratory will proceed as normal and all experiments performed therein will function properly.  However, in a frame of reference outside of the CTC, an inertial observer may see such bizarre occurrences as shards of glass collecting themselves in to a beaker and rising in to the air.  If the laboratory is caused to move into and out of the CTC, an experimenter inside the lab may find herself arriving before she left, as measured by exterior clocks, even if her own clocks show an increase in time \cite{Janca}.

While this violation of causality is allowable in the classical understanding of GR, Stephen Hawking has proposed the Chronology Protection Conjecture in order to, as he says,``make the universe safe for historians'' \cite{Hawking}.  In his paper presenting this conjecture, Hawking claims that any curved spacetime allowing CTCs is impossible to construct in GR because it will require unphysical distributions of stress-energy, such as negative energy density.  To go further, Hawking then demonstrates that even if such a spacetime were to exist, it would be impossible for an object to move in to the region containing CTCs due to divergences in its quantum mechanical propagator and the resulting re-curvature of spacetime from the energy needed to reach that region.  Kim and Thorne \cite{KimThorne} likewise find this problem of diverging stress-energy, but assert that perhaps quantum gravitation effects will limit it; whether this is or is not the case will depend on the form of the eventual theory of quantum gravitation, which is still being debated.  In this present work, we are not dealing with actual gravity, and hence quantum gravitational effects are not of interest.  Further, we do not need to approach the non-causal region directly; as mentioned earlier, we could also use fiber optic cables, completely bypassing this objection.

In pursuit of chronology protection, Hawking briefly considers a metric very similar to our own, namely
$$ds^2 = -dt^2+dr^2+ r^2(d\theta^2+\sin^2\theta d\phi^2)-fd\phi^2+2fd\phi dt ,$$
for $f=r^2t^2\sin^4\theta\sin^2\frac{\pi r}{a}$, to demonstrate the inability of causality violation.  This spacetime differs from ours mostly in the form of the frame-dragging coefficients.  Hawking's principal objection to such a spacetime is that, based on the Einstein equations, such a metric must come from negative energy density; otherwise, Hawking accepts that such a spacetime will in fact lead to violations of causality if objects could reach the non-causal regions. This issue of negative energy density is not a problem for us in TO, since we intend to construct our optical spacetime from a metamaterial, which does not have to follow any sort of energy condition.  Therefore, while Hawking's conjecture offers a strong argument against general relativistic time machines to the past, neither of his arguments are of interest to the Mirror.

From a general relativistic point of view, light moving in a spacetime with a metric such as \eqref{themetric} will violate causality.  While effects of GR are predicted to stop this, these effects have no bearing in transformation optics.  Therefore, if the material can be built, then theory predicts that it can be used to violate causality.  Since a violation of causality has never been observed, we should look closer at the material.

Looking at the MPs in \eqref{galadrieltensors}, they are seen to be anisotropic, much like other systems studied with TO.  Such anisotropies are also present in the invisibility cloak, which has been subject to much study and lately constructed in reduced models \cite{TheCloak}.  These anisotropies are usually achieved by means of metamaterials, such as those made of lattices of split-ring resonators.   Each lattice site can be modeled as an RLC circuit, where the loop of the ring itself is the inductor and the tiny gap in the ring serves as the capacitor.  From this, we find the permittivity and permeability follow a Lorentz model \cite{Magnetomaterial}.  If we put $\epsilon$ in to cylindrical components, it is diagonal with $\epsilon_r = K/r, \epsilon_\phi = r/K,$ and $\epsilon_z = K/r$.  In the Lorentz model, if we look at specifically the $\phi$-component of $\epsilon$, we have \cite{Jackson}
\begin{equation}\label{Lorentz}
	\epsilon_\phi(\omega) = 1 - \frac{\omega_P^2}{\omega_o^2-\omega^2-i\omega\gamma},
\end{equation}
where $\omega_P$ is a function of the capacitance and inductance, $\omega_o$ is the resonant frequency of the circuit, and $\gamma$ is related to the resistivity, specifying the loss in the circuit.

Now consider what happens as light moves in such a material.  For simplicity, we will look only at light in an isotropic medium that follows the Lorentz model.  There, we have
\begin{equation}\label{illustration1}
	\vec{D}(t) = \frac{1}{\sqrt{2\pi}} \int_{-\infty}^\infty d\omega \; \epsilon(\omega) \vec{E}(\omega) e^{-i\omega t}.
\end{equation}
As is well-known, if we write $\vec{E}(\omega)$ as a Fourier transform of the time domain and insert into \eqref{illustration1}, and exchange the order of integration, the above can be simplified to
\begin{equation}
	\vec{D}(t) = \vec{E}(t) + \int_{-\infty}^\infty d\tau \; \vec{E}(t-\tau) \chi(\tau)
\end{equation}
where $\chi(\tau)$ is a response function telling us how strongly the electric field $\vec{E}(t-\tau) $ at time $t-\tau$ contributes to the field $\vec{E}(t)$ at the present.  Notice, for $\tau < 0$, it is not specifically ruled-out that the future fields contribute to the present fields; if $\chi(\tau)$ is non-zero for $\tau<0$, then we will have the future electric fields influencing the present.  However, for the Lorentz model, the response function takes the form of a contour integral,
\begin{equation}\label{response}
	\chi(\tau) = \frac{1}{2\pi} \int_{-\infty}^\infty d\omega \; \frac{\omega_P^2 e^{-i\omega \tau}}{\omega_o^2 - \omega^2 - i\gamma\omega}.
\end{equation}
This integrand has poles at $\omega =- i\frac{\gamma}{2} \pm\sqrt{\omega_o^2 - \frac{\gamma^2}{4}} ,$ which are in the lower half-plane.  When $\tau > 0$ (meaning we are considering past contributions), our contour is in the bottom half-plane, and we pick up contributions from these two poles.  However, when $\tau < 0$ (meaning we are considering future contributions), our contour is in the top half-plane, which has no zeroes, so $\chi(\tau) = 0$ for all $\tau <0$.  Therefore, in the Lorentz model, future fields do not affect the present \cite{Jackson}.  Though we have applied this thought to a much simpler situation than that of the bi-ansitropic material in the Mirror, it still serves to illustrate the problem.

We further notice that this effect is due to a non-zero resistivity $\gamma$; for if $\gamma=0$ above, then the poles of the contour integral in \eqref{response} fall on the real axis, which will contribute for both half-planes (that is, from the future and the past).  This is a curious point.  The resistivity $\gamma$ represents in a certain sense the loss of energy in the system due to heat as current passes through a resistive element.  This suggests entropy.

We can then see a more general principle.  Any real material -- Lorentzian or not -- as electric fields move through it, must undergo loss to heat.  Imagine, then, light within the Mirror from the future event $t_n$ moving to a slightly earlier event $t_{n-1}$ as it travels along a CNG.  To do this, the electric fields of the light must interact with the elements, which produces some amount of heat $\delta Q$ over the time $t_{n-1}-t_{n}$, which gets transmitted as waste heat to the air in the room.  In the lab frame, this means that the overall heat of the system is \emph{decreasing} over time $t_n-t_{n-1}$, and that waste heat $\delta Q$ is being absorbed \emph{by} the element \emph{from} the air to go in to diverting the path of light.  This clearly entails a decrease in entropy in a process that does nothing more than convert heat to useful work (the work in the response of the material to the light fields).  Hence, for any real material, no matter what its material parameters may do on paper, it is impossible for the future to communicate with the past.   We find this consideration the strongest, as the Second Law has thus far proven unassailable.

%%%%%%%%%%%%%%%%%%%%%%%%%%%%%%%%%%%%%%%%%%%%%%%%%%%
%			CONCLUSION
%%%%%%%%%%%%%%%%%%%%%%%%%%%%%%%%%%%%%%%%%%%%%%%%%%%
\section{Conclusions and Future Work\label{Conclusions}}

We have seen then that the theoretical framework behind transformation optics allows for materials that support closed null geodesics.  We have explicitly constructed just such a material ourselves.  These materials, if possible to construct, have the predicted effect of allowing for the violation of causality, with information from the future being able to reach the past.  This behavior follows from nothing more than the macroscopic material properties of the medium, in terms of how it responds to incoming E\&M fields.

However, it is ultimately the microscopic behavior of the medium that steps in to prohibit the Mirror from working.  This raises an interesting limitation to transformation optics, which usually works only with the relations for the macroscopic material parameters in terms of the metric tensor.  This suggests the need for a set of equations -- analogous to the energy conditions from general relativity -- that can relate the metric tensor to the microscopic material parameters, which have definite thermodynamic limits.  Such equations would be able to quickly tell researchers the limits of their transformation media.

In conclusion, thermodynamics will prevent the Mirror from violating causality.   Thus, even for transformation optics, history will indeed remain safe for historians.

\begin{acknowledgments}
Acknowledgments are made to the Center for Simulational Physics at UGA, and especially to Steve Lewis and Bill Dennis for their support.  Special thanks to Tomas Koci at the CSP for furious and enlightening debate about this paper.  Additional thanks to Ben Jackson, Amara Katabarwa, HB Schuttler, and Andrei Galiautdinov  for reviewing early drafts of this paper.  My deepest gratitude to Angela LeClair for providing the graphics in Figure \ref{aliceandbob}.
\end{acknowledgments}

%%%%%%%%%%%%%%%%%%%%%%%%%%%%%%%%%%%%%%%%%%%%%%%%%%%
%			REFERENCES
%%%%%%%%%%%%%%%%%%%%%%%%%%%%%%%%%%%%%%%%%%%%%%%%%%%
\bibliography{references.bib}

\end{document}